\documentclass[final,5p,times,twocolumn]{elsarticle}

\usepackage{amssymb}
\journal{Results in Physics}

\begin{document}

\begin{frontmatter}



\title{Inertial-Hall effect: the influence of rotation on the Hall conductivity}

\author{Julio E. Brand\~ao and F. Moraes} 
\address{ Departamento de F\'{\i}sica, CCEN,  Universidade Federal 
da Para\'{\i}ba, Caixa Postal 5008, 58051-970 , Jo\~ao Pessoa, PB, Brazil }
\author{M. M. Cunha}
\address{Departamento de F\'{\i}sica, CCEN, Universidade Federal de Pernambuco, 50670-901 Recife, PE, Brazil}
\author{Jonas R. F. Lima}
\address{Instituto de Ciencia de Materiales de Madrid (CSIC) - Cantoblanco, Madrid 28049, Spain}
\ead{jonas.iasd@gmail.com}
\author{C. Filgueiras}
\address{Unidade Acadmica de F\'{\i}sica, Universidade Federal
de Campina Grande, POB 10071 Campina Grande, PB 58109-970, Brazil}

\begin{abstract}
Inertial effects play an important role in classical mechanics but have been largely overlooked in quantum mechanics. Nevertheless, the analogy between inertial forces on mass particles and electromagnetic forces on charged particles is not new.  In this paper, we consider a rotating non-interacting planar two-dimensional electron gas with a perpendicular uniform magnetic field and investigate the effects of the rotation in the Hall conductivity. The rotation introduces a shift and a split in the Landau levels. As a consequence of the break of the degeneracy, the counting of the states fully occupied below the Fermi energy increases, tunning the Hall quantization steps. The rotation also changes the quantum Hall plateau widths. Additionally, we find the Hall quantization steps as a function of rotation at a fixed value of the magnetic field.
\end{abstract}


\end{frontmatter}


\section{Introduction}

Since the discovery of the integer quantum Hall effect (IQHE) in 1980 by von Klitzing \textit{et. al.} \cite{Klitzing}, the two-dimensional electron gas (2DEG) in a strong perpendicular magnetic field has been a subject of intense study, both experimentally and theoretically. The IQHE is a macroscopic effect of solid state physics and it is characterized by a quantized Hall conductivity which is given by integer multiples of $e^2/h$, where $e$ is the electrical charge and $h$ is the Planck\rq{}s constant. The quantization of the Hall conductivity has been measured to 1 part in $10^9$ \cite{Poirier,SCHOPFER}. This precision reveals the topological nature of the Hall conductivity, which does not depend on the material, geometry and microscopic details of the sample, and makes the IQHE very useful in the field of metrology. The properties of a charged particle in a magnetic field are important also in other fields as high energy physics, atomic physics and astrophysics, as was pointed out in \cite{Klitzing15092005}, which has attracted even more interest in the study of the IQHE.

The Coriolis force acts on a particle of mass $m$ very much like the magnetic force on a charged particle. This analogy has been explored by Aharonov and Carmi \cite{springerlink:10.1007/BF00709020,springerlink:10.1007/BF00709117} in the early 1970's, by Sakurai \cite{PhysRevD.21.2993} in 1980 and by Tsai and Neilson \cite{PhysRevA.37.619} in 1988  in the context of a rotational quantum phase similar to Aharanov-Bohm's. The idea of rotation working as an effective magnetic field is in fact  quite old. In 1915 Barnett \cite{PhysRev.6.239,RevModPhys.7.129} already published a paper on magnetization by rotation which has recently had a renewed interest  applied to nanostructures \cite{bretzel:122504,PhysRevB.81.024427}. A rotational analog of the classical Hall effect has been proposed \cite{springerlink:10.1023/A:1021043923114} and the inertial effects of rotation in spintronics studied \cite{PhysRevLett.106.076601,PhysRevB.84.104410,matsuo:242501}. Based on the same analogy, Dattoli and Quattromini \cite{2010arXiv1009.3788D}, introduced Coriolis quantum states analogous to Landau levels (LLs).  This analogy also appears in the study of rapidly rotating Bose-Einstein Condensates \cite{0953-8984-20-12-123202}, for the Hamiltonian describing a rotating gas in a harmonic trap is similar to that for charged particles in a magnetic field. The subject of analogue Landau levels has been recently approached in the more general context of combined non-inertial, gravitational and electromagnetic effects by Konno and Takahashi \cite{PhysRevD.85.061502} who were interested on quantum states on the surface of a rotating star. The Quantum Hall effect under rotation has been discussed in more general grounds in \cite{johnson:649,0295-5075-54-4-502}.
 
The Coriolis force does not come alone. Its companion, the centrifugal force, will be also felt by the particle in the rotating system. Together, the Coriolis and centrifugal contributions to the quantum Hamiltonian lead only to a coupling between the particle angular momentum and the rotation, for the case of a spinless particle. This gives rise to non-degenerate, sample-length dependent, Landau levels \cite{johnson:649,0295-5075-54-4-502}.  Neglecting the centrifugal part, besides this coupling, there appears the richer structure of rotational Landau levels \cite{2010arXiv1009.3788D}. On the other hand, if we are free from the centrifugal force we end up with a Landau levels system that includes the coupling between the particle angular momentum and the rotation. It should be notice that, if a steady time variation of the rotation is assumed, then the Euler force should be included in the analysis.  

The system of our interest consists in a non-interacting free electron gas in a rotating planar conductor with a uniform magnetic field applied perpendicular to the rotating plane. Our purpose is to investigate the quantum Hall effect in this system, analyzing the influence of the rotation in the Hall conductivity. Charged particles in a rotating Hall sample were already studied in Ref. \cite{0295-5075-54-4-502}, where it was pointed out that the quantization of the Hall conductivity is not affected by the rotation. However, as it will be shown here, the rotation breaks same degeneracy of the LLs and the counting of states fully occupied bellow the Fermi energy may change, altering the Hall quantization steps.

The paper is organized as follows. In Sec. 2 we write out the Hamiltonian of a charged particle in a rotating disk in the presence of a magnetic field and find the energy spectrum. In Sec. 3 we analyze the electronic structure, showing that the rotation induces a shift and a split in the Landau levels. In Sec. 4 we investigate the influence of the rotation in the Hall conductivity. The paper is summarized and concluded in Sec. 5.

\section{The spinless charged particle}

\begin{figure}[!htb]
\includegraphics[scale=0.4]{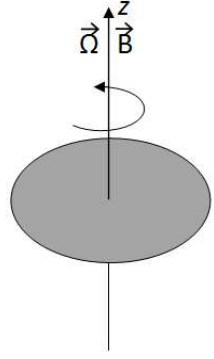}
\label{degenerecenciap=2}
\caption{A rotating disk with a perpendicular uniform magnetic field. The rotation speed and magnetic field vector are in the $z$ direction.}
\label{disco}
\end{figure}

Let us consider a free particle in a rotating disk with a uniform magnetic field perpendicular to the disk [see Fig. \ref{disco}]. The Coriolis and centrifugal forces are given by
\begin{equation}
\vec{F}_{Cor}=2m(\vec{v}\times \vec{\Omega}) , 
\end{equation}
and
\begin{equation}
\vec{F}_{Cen}=-m \vec{\Omega}\times (\vec{\Omega}\times \vec{r}) ,
\end{equation}
respectively. These forces enter the Schr\"odinger Hamiltonian as a vector and scalar inertial potential \cite{springerlink:10.1007/BF00709020,springerlink:10.1007/BF00709117} given by
\begin{equation}
\vec{A}_{ine}=\frac{1}{2}(\vec{\Omega} \times \vec{r})
\end{equation}
and
\begin{equation}
V_{ine}=-\frac{1}{2}(\vec{\Omega} \times \vec{r})^2 \; ,
\end{equation}
respectively, and the Hamiltonian is written as
\begin{equation}
H=\frac{1}{2m}(\vec{p}-2m\vec{A}_{ine})^2+mV_{ine} \; .
\end{equation}
A magnetic field $\vec{B}$ applied in the laboratory will be felt by charged particles in the rotating reference frame as an electric and a magnetic field given by \cite{PhysRevB.84.104410}
\begin{equation}
\vec{E}^{\prime} = (\vec{\Omega} \times \vec{r})\times \vec{B}
\end{equation}
and
\begin{equation}
\vec{B}^{\prime} = \vec{B} \; .
\end{equation}

Therefore, the Hamiltonian in cylindrical coordinates of a particle in a rotating disk in the presence of a magnetic field, with $\vec{\Omega}=\Omega  \hat{z}$ and $\vec{B}=B \hat{z}$, can be written as
\begin{eqnarray}
H=\frac{[\vec{p}-q\vec{A}-m(\vec{\Omega}\times\vec{r})]^2}{2m}-\frac{m(\vec{\Omega}\times\vec{r})^2}{2}+qV\; ,\label{hamiltonianageral}
\end{eqnarray}
where $V$ and $\vec{A}$ are the scalar and vector electromagnetic potentials, and are given by
\begin{eqnarray}
V=-\frac{\Omega Br^2}{2},\\
\vec{A}=(0,\frac{Br}{2},0).
\end{eqnarray}
Thus, the Hamiltonian can be summarized to
\begin{eqnarray}
H=\frac{p^2}{2m}-\alpha rp_\phi+\beta r^2 \; ,
\end{eqnarray}
with
\begin{eqnarray}
\alpha=\frac{qB}{2m}+\Omega ,\\
\beta=\frac{q^2B^2}{8m}.
\end{eqnarray}
For this Hamiltonian, the Schr\"odinger equation can be written as
\begin{eqnarray}
-\frac{\hbar^2}{2m}\nabla^2\psi+i\alpha\hbar\frac{\partial\psi}{\partial\phi}+\beta r^2\psi=E\psi . \label{eqschrodinger}
\end{eqnarray}
With the \textit{ansatz} $\psi=R(r)e^{-il\phi}$ , Eq. (\ref{eqschrodinger}) becomes
\begin{eqnarray}
r^2R''+rR'+(-\sigma^2r^4+\lambda r^2-l^2)R=0, \label{eqemr}
\end{eqnarray}
where $\sigma^2=\frac{q^2B^2}{4\hbar^2}$ and $\lambda=\frac{2m}{\hbar}\left(\frac{E}{\hbar}-\frac{qBl}{2m}-\Omega l\right)$. Writing $\sigma r^2=\xi $ and looking to the asymptotic limits when $\xi\rightarrow\infty$ and $\xi\rightarrow 0$, one can propose a solution of the form
\begin{eqnarray}
R(\xi)=e^{-\frac{\xi}{2}}\xi^{\frac{|l|}{2}}u(\xi) \;.
\end{eqnarray}
Replacing it in Eq. (\ref{eqemr}), one gets
\begin{eqnarray}
\xi\frac{d^2u}{d\xi^2}+\left[1+|l|-\xi \right] \frac{du}{d\xi}+\left[ \frac{\lambda}{4\sigma}-\frac{1}{2}\left(|l|+1 \right) \right]u=0 \; ,\label{hiperour}
\end{eqnarray}
that is a Confluent Hypergeometric Equation, which has solution
\begin{eqnarray}
u=A \cdot F\left(\frac{-\lambda}{4\sigma}+\frac{1}{2}\left(|l|+1 \right),1+|l|,\xi \right),
\end{eqnarray}
where A is a constant and $F(a,b,z)$ is a confluent hypergeometric function, in this case, degenerate. In order to have a finite polynomial function (the hypergeometric series has to be convergent in order to have a physical solution), the condition $a=-n$ has to be satisfied, where $n$ is a positive integer number. From this condition, the discrete possibles values for the energy are given by
\begin{eqnarray}
E_{n,\ell}=\hbar \omega_c \left\lbrace n+\frac{\ell}{2} +\frac{|\ell|}{2}+\frac{1}{2}\right\rbrace +\Omega l\hbar\; ,\label{energia1}
\end{eqnarray} 
where $\omega_c=qB/m$ is the cyclotron frequency. 
The wave function is
\begin{eqnarray}
\psi&=& A e^{-il\phi-\frac{\sigma r^2}{2}}(\sigma r^2)^{\frac{|l|}{2}} \nonumber \\
&&\times  F\left(-\frac{\lambda}{4\sigma}+\frac{1}{2}\left(|l|+1 \right),1+|l|,\sigma r^2\right).\label{fdeonda}
\end{eqnarray}
One can notice that the values of $\vec{B}$ and $\vec{\Omega}$ are arbitrary. Therefore, one can adjust the rotation and the magnetic field for different values. 

Considering $\Omega=0$ \cite{landau}, there is only the magnetic force acting and, therefore, we find the usual Landau levels in (\ref{energia1}),
\begin{eqnarray}
E_{n}=\hbar\omega_c\left( n+\frac{\ell}{2}+\frac{|\ell|}{2}+\frac{1}{2}\right)=\hbar\omega_c\left(m+\frac{1}{2}\right).
\end{eqnarray}

With $B=0$,  we have only inertial forces acting in the system. This was done in 1999 by B.L. Jonhson \cite{johnson:649}, who showed that
\begin{eqnarray}
\psi&=&J_\ell(|\lambda|r)e^{-i\ell\phi},
\end{eqnarray}
and
\begin{eqnarray}
E&=&\frac{\hbar^2\lambda^2}{2m}+\ell\hbar\Omega, \label{energiajohnson}
\end{eqnarray}
where $J_{\ell}$ is the Bessel function and the energy spectrum is obteined from the boundary conditions in the disk. One can find this result calculating the limit of the function (\ref{fdeonda}) when $B\rightarrow0$, and extracting from the new wave function (Bessel), the new condition for the energy. It's important to note that one cannot obtain the expression (\ref{energiajohnson}) only substituting $B=0$ in our energy spectrum (\ref{energia1}), because the energy must be obtained using the boundary conditions that is provided by the Schr\"odinger equation. So, when we choose $B=0$ in the Schr\"odinger equation, the Bessel equation is obtained, no longer the hypergeometric equation. Note that for $B\neq0$, the resulting equation is hypergeometric, resulting in a Landau-like spectrum that is independent of the edge of the sample. It is a consequence of the fact that a weak magnetic field is enough to confine the wave function, so the edge is not important. However, with $B=0$, there are quantized non-degenerated energy levels that are influenced by the edge of the sample.

\section{Electronic structure}

Before analyzing the electronic structure for a general case, let us first consider two special choices for the rotation and the magnetic field. The first case is $\Omega=-\frac{qB}{m}$, which represents a system that has only magnetic and Coriolis forces, \textit{i.e.}, the electric and centrifugal forces vanish. The second one is $\Omega=-\frac{qB}{2m}$, which represents the inverse. The magnetic and Coriolis forces vanish, remaining only the electric and centrifugal forces. Then, substituting these choices in Eq. (\ref{energia}), one can find that the energy levels are given by
\begin{eqnarray}
E_{n'}=\hbar\omega_c\left( n'+\frac{1}{2}\right), \label{energyfirstcase}
\end{eqnarray}
for the first case and 
\begin{eqnarray}
E_{n''}=\hbar\omega_c \left( n''+\frac{1}{2}\right), \label{energysecondcase}
\end{eqnarray}
for the second case, where $n'=n+\frac{|\ell|}{2}-\frac{\ell}{2}$ and $n''=n+\frac{|\ell|}{2}$. One can note that in the second case, even though with the vanishing of the  magnetic force and without the vector potential appearing in the final Hamiltonian, a Landau-like quantization still exists.

\begin{figure}[!htb]
\includegraphics[scale=0.4]{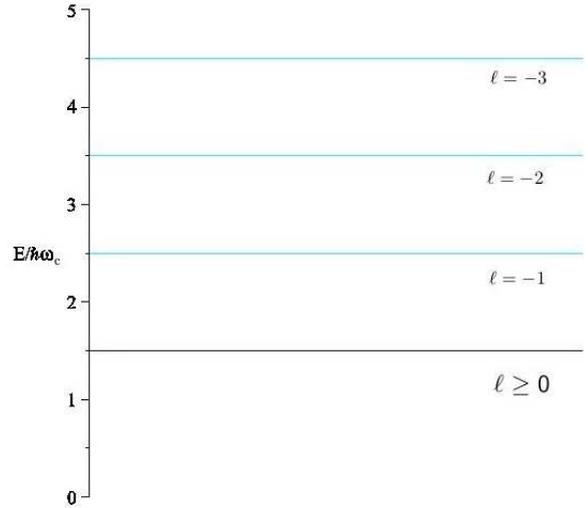}
\label{degenerecenciap=2}
\caption{The energy levels obtained from the energy spectrum (\ref{energyfirstcase}) for $n=1$. All the levels with $\ell \geq 0$ are degenerate}
\label{E1}
\end{figure}

\begin{figure}[!htb]
\includegraphics[scale=0.4]{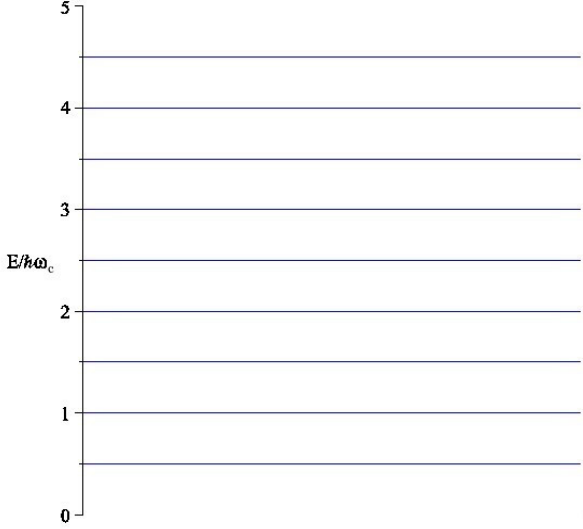}
\label{degenerecenciap=1}
\caption{The energy levels obtained from the energy spectrum (\ref{energysecondcase}) for $n=1$. The possible values for the energy are separated by $\hbar \omega_c/2$, as in the harmonic oscillator.}
\label{E2}
\end{figure}

The states are degenerate in both cases. One can see this by analyzing the numbers $n'$ and $n''$ in Eqs. (\ref{energyfirstcase}) and (\ref{energysecondcase}). In the first case, different combinations between $n$ and $\ell$ give the same $n'$, which means the same energy. In particular, all states with $\ell\geq 0$ for a given $n$ are degenerate [see Fig. \ref{E1}]. This is what happens in the usual Landau levels, except that in the LLs for the same value of $n$, the states are infinitely degenerated with $\ell\leq 0$. So, the result of the first case is the usual Landau levels with reversed charge. It's important to note that  the number $n'$ is always integer for any combination between $n$ and $\ell$. In the second case, it is also possible to obtain different combinations between $n$ and $\ell$ resulting in the same value for $n''$. One can note that $n''$ is integer or half-integer, which means that the energy gap between two levels will be $\hbar\omega_c/2$ [see Fig. \ref{E2}]. All levels are infinitely degenerated too. Then, in the second case the energy levels are equivalent to the ones of the harmonic oscillator.

\begin{figure}[!htb]
\includegraphics[scale=0.55]{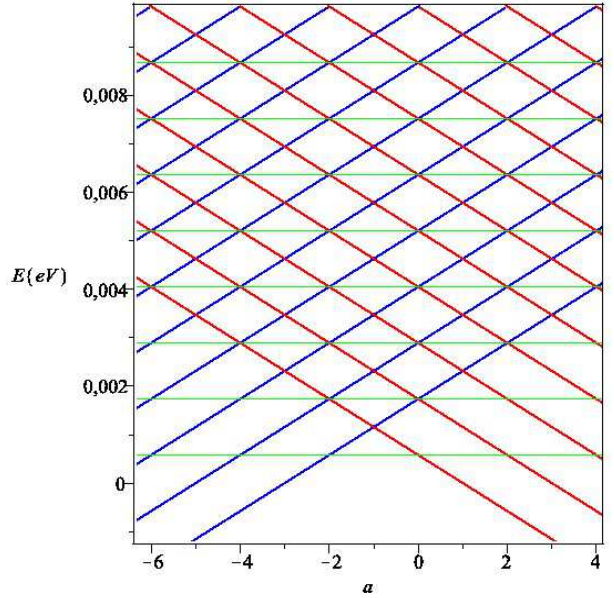}
\label{energiaxrotacaozoom}
\caption{The energy levels obtained from the energy spectrum (\ref{energia}) for $\ell=-1$ (red), $\ell=0$ (green) and $\ell=1$ (blue) as a function of the parameter $a$. Each line represents a value of $n$ and the lowest line is $n=0$. When $a$ increases, the energy levels with positive angular momentum are shifted up, states with negative angular momentum are shifted down and the levels with null angular momentum stay unmoved.}
\label{Ea2}
\end{figure}

In order to consider the general case one can relate the rotation and the magnetic field as
\begin{equation}
\Omega=\frac{a}{2}\frac{qB}{m}, \label{a}
\end{equation}
where $a$ is a real number. This expression cover all possible combinations between $B$ and $\Omega$. One can see that the two special cases discussed previously represent $a=-2$ and $a=-1$, respectively. So, the energy levels can be written as 
\begin{eqnarray}
E_{n,\ell}&=&\hbar\omega_c\left\lbrace \left(\frac{a}{2}+\frac{1}{2}\right)\ell  +\frac{|\ell|}{2}+n+\frac{1}{2}\right\rbrace  \; ,
\label{energia}
\end{eqnarray}
which allows us to plot the energy spectrum as a function of $a$, as done in Figs. \ref{Ea2}. The red, green and blue lines represent states with $\ell=-1,0,1$ respectively. Each line represents a value of $n$, and the lowest line is $n=0$. As can be seen, the levels with positive angular momentum are shifted up when $a$ increase, while states with negative angular momentum are shifted down. The levels with null angular momentum do not change with $a$. One can see that the states with $\ell=1$ and $\ell=-1$ are symmetric with relation to $a=-1$, which explains the fact that the special case with $a=-2$ is equivalent to the usual LLs $(a=0)$ with reversed charge. In consequence of this symmetry, there are crossings between states with $\ell=1$ and $\ell=-1$ at $a=-1$, which mean that they are degenerate. The crossings remain at other integer values of $a$, but in these cases the degenerescence is between states with different values of $n$. When $a$ is an even integer the levels with $\ell=0$ are degenerate with the states with $\ell=-1$ and $\ell=1$, which does not happen at an odd integer value of $a$.  There is no degenerescence among these states when $a$ is not an integer.

Therefore, as was discussed above, the rotation affects directly the Landau levels, inducing a shift and a split of the states. The shift has no important physical consequences, however the split of the levels will affect, for instance, the quantum Hall effect. This will be discussed in the next section.

\section{Hall conductivity}

At zero temperature and considering that the Fermi energy $E_F$ is in an energy gap, the Hall conductivity can be written as \cite{Streda}
\begin{equation}
\sigma_H(E_F,0)=\frac{e}{S}\frac{\partial N}{\partial B} \; ,
\end{equation}
where $N$ is the number of states below the Fermi energy and $S$ is the area of the surface. The density of states is given by
\begin{equation}
n(E)=\frac{|eB|}{2\pi \hbar}\sum_{n,l}\delta(E-E_{n,l}) \; . 
\end{equation}
Therefore, one can obtain $N$, that is given by
\begin{equation}
N=S\int_{-\infty}^{E_F}n(E)dE=\frac{S|eB|}{h} \times n
\end{equation}
where $n$ is the number of fully occupied LLs below $E_F$. Thus, the Hall conductivity is
\begin{equation}
\sigma_H=-\frac{e^2}{h}n \; .
\end{equation}
One can notice that the Hall conductivity obtained here has the same form as in the case without rotation, in agreement with \cite{0295-5075-54-4-502}. Nevertheless, due to the break of the degeneracy of the LLs induced by rotation, the number of states fully occupied below the Fermi energy may change.

\begin{figure}[!htb]
\includegraphics[scale=0.75]{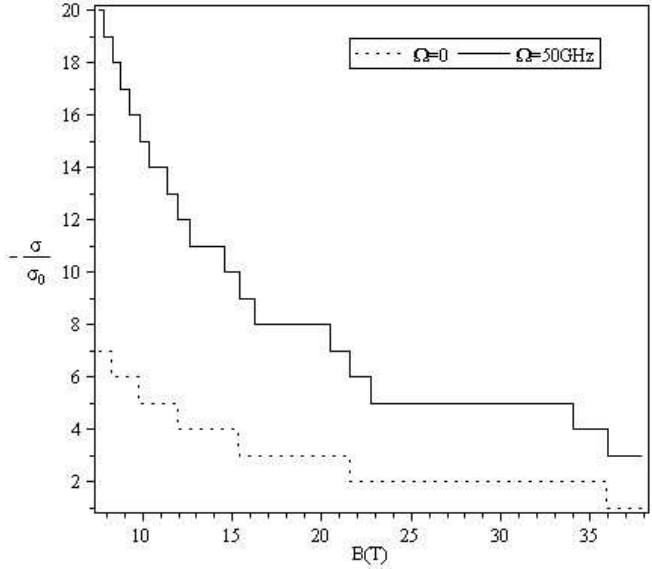}
\label{degrausomegafix}
\caption{The Hall quantization steps as a function of the magnetic field for $\Omega=0$ (dotted line) and $\Omega=50$ GHz (continuum line). We are considering only states with $\ell=-1,0,1$ and $E_F=6.24$ meV. Each step for case with $\Omega=0$ becomes three when $\Omega \neq 0$.}
\label{hallb}
\end{figure}

In Fig. \ref{hallb} we plot the Hall conductivity as a function of the magnetic field for $\Omega=0$ and $\Omega=50$ GHz. In fact, it is plotted $-\sigma_H/\sigma_0$, which is the number of fully occupied LLs below $E_F$, where $\sigma_0=e^2/h$. When the magnetic field increases, all states increase their energy and begin to cross the Fermi Level, creating the Hall steps. It should be mentioned that we are considering only states with $\ell =-1,0,1$. In the case with $\Omega=0$, when the magnetic field increases, the next step is always wider than the previous one. This is a consequence of the fact that the distance between two subsequent Landau levels increases with the magnetic field. In contrast, for $\Omega \neq 0$, the next step is not necessarily wider than the previous one. It is due to the break of the degeneracy among states with different values of $\ell$ introduced by rotation. So, each step, in the case with $\Omega=0$, becomes three for $\Omega \neq 0$. Therefore, for $\Omega \neq 0$, each step is wider than the previous third.

\begin{figure}[!htb]
\includegraphics[scale=0.7]{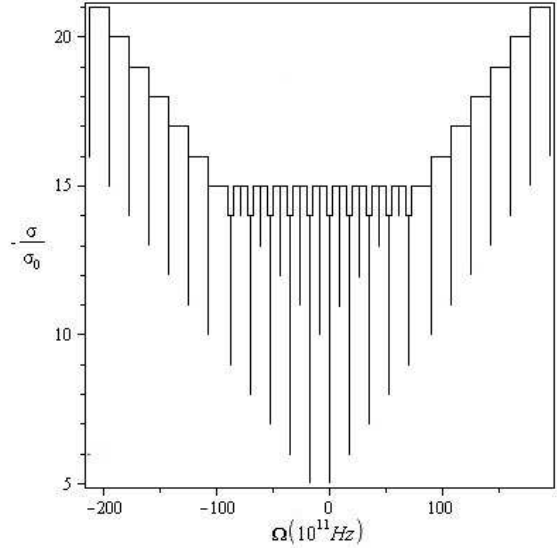}
\label{degrausBfix}
\caption{The Hall quantization steps as a function of the rotation for $B=10$ T and $E_F=6.24$ meV. When the rotation  modulus is lower than $100\times 10^{11}$ Hz the Hall conduvtivity oscillates. For higher values of the rotation speed our model is not valid.}
\label{hallr}
\end{figure}

The Hall conductivity versus the rotation speed with a fixed value of the magnetic field is plotted in Fig. \ref{hallr}. Looking to the energy spectrum (\ref{energia1}) we see that when the rotation speed increases, states with positive values of $\ell$ are shifted up, while states with negative $\ell$ are shifted down. Energy levels with null angular momentum do not change with the rotation. Thus, changing the rotation there will be states crossing up and down the Fermi energy, enabling different behaviors for the Hall quantization steps. In Fig. \ref{hallr} we are considering only states with $\ell = -1,0,1$, as in the previous case. It is possible to see that the number of states fully occupied below the Fermi energy can increase, decrease or oscillate when the rotation changes continuously. The negative values of $\Omega$ in Fig. \ref{hallr} means that the rotation is in an opposite direction with relation to the magnetic field.

At specific values of $\Omega$ there are a sharp drop in the conductivity, which are represented by the vertical lines in Fig. \ref{hallr}. It happens because for these values of $\Omega$ the parameter $a$ is an integer and states with $\ell = -1,1$ are degenerate, as can be seen in Fig. \ref{Ea2}, which reduce the number of states below the Fermi level. Then, its possible to understand Fig. 6 looking with attention to Fig. 4. When the modulus of the rotation speed increases up to values around $100 \times 10^{11}$ Hz, the Hall steps have a teeth-like aspect, independent of the direction of the rotation, which means that after an energy level crosses up (down) the Fermi energy another level crosses it down (up). For higher values of the modulus of the rotation speed the number of fully occupied levels below $E_F$ begins to increase, in contrast to the previous case, where the Hall conductivity decreases when the magnetic field increases.

It is important to say that for high values of the rotation speed our model is not valid, because here we are not considering relativistic effects. The velocity $v$ in the edge of the sample is given by $v=R\Omega$, where $R$ is the radius of the disk. So, in order to have no relativistic values of $v$, the radius of the disk where our model in valid decreases when $\Omega$ increases.
% So, when the modulus of the rotation increases up to values around $100 \times 10^{11}$ Hz, the Hall steps have a teeth-like aspect, independent of the direction of the rotation, which means that after an energy level crosses up (down) the Fermi energy another level crosses it down (up).  For higher values of the modulus of the rotation speed the number of fully occupied levels below $E_F$ begins to increase, in contrast to the previous case, where the Hall conductivity decreases when the magnetic field increases.

%Errado! It is not possible to see in Fig. \ref{hallr}, but the Hall conductivity changes only at the values of $\Omega$ corresponding to $a$ integer. This is a consequence of the existing degenerescence among the levels when $a$ is an integer.

\section{Conclusion}

In this paper, we investigated the influence of the rotation on the quantized Hall conductivity. For this purpose, we considered a rotating non-interacting planar two-dimensinal electron gas with a perpendicular uniform magnetic field. We verified that the rotation breaks the degenerecence of the Landau levels. However, when the relation $\Omega=\frac{a}{2}\frac{qB}{m}$ is satisfied and $a$ is an integer, there are degenerate states. We analyzed the electronic structure for all possible values of the parameter $a$ and emphasized two special cases, $a=-2$ and $a=-1$. It was shown that at $a=-2$ the electronic structure is equivalent to the usual Landau levels, but with reversed charge. Whereas, at $a=-1$ the energy levels are equivalent to the ones of a harmonic oscilator. We found the Hall conductivity, which is the same as in the case without rotation, however, due to the split introduced by rotation, the counting of the states fully occupied below the Fermi energy change. Fixing the rotation speed and plotting the Hall conductivity as a function of the magnetic field for states with $\ell=-1,0,1$, we found that each Hall step for the case with $\Omega=0$ becomes three when $\Omega \neq 0$, increasing the height of the Hall quantization steps. Additionally, we plotted the Hall conductivity against the rotation at a fixed value of the magnetic field. It was shown that the Hall conductivity oscillates when the rotation increases up to values around $100 \times 10^{11}$ Hz, for the values of the parameters chosen here. This happens because states with positive (negative) angular momentum are shifted up (down) when the rotation speed increases. As a perspective, one can analyze the influence of rotation in the quantum Hall effect of novel materials such as graphene and topological insulators.

{\bf Acknowledgements}: We are grateful to CNPq, CNPq-MICINN bilateral and CAPES for financial support. J. R. F. Lima thanks the hospitality of the ICMM where part of this work was developed.


%% TeX f
\end{document}